\documentclass[aps,prb,twocolumn,footinbib]{revtex4-2}
\usepackage[a4paper,top=1.5cm,bottom=1.5cm,left=1.75cm,right=1.75cm]{geometry}
\usepackage{amsmath}
\usepackage{amssymb}
\usepackage{enumerate}
\usepackage{color}
\usepackage{mathptmx}
\usepackage{mathabx} % needed for widehat
\usepackage{epsf}
\usepackage{graphicx}
\DeclareGraphicsExtensions{.eps, .pdf, .jpg, .png}
\usepackage{hyperref}
\def\nicefrac#1#2{\genfrac{}{}{}{1}{#1}{#2}}
\newcommand{\be}{\begin{equation}}
\newcommand{\ee}{\end{equation}}
\newcommand{\ber}{\begin{eqnarray}}
\newcommand{\eer}{\end{eqnarray}}
\newcommand{\kb}{k_{\text{B}}}
\def\ns{\negthickspace}
\def\Tr#1{\mbox{Tr}\big\{#1\big\}}
\def\tz{{\widehat{\tau}_3}}
\newcommand{\tauz}{{\tau}_3}
\newcommand{\whmfGk}{\widehat{\mathfrak{G}}^{\text{K}}}
\def\whDelta{\widehat{\Delta}}
\def\cA{{\mathcal A}}
\def\cY{{\mathcal Y}}
\def\vj{{\bf j}} 
\def\vn{{\bf n}} 
\def\vp{{\bf p}} 
\def\vq{{\bf q}} 
 
\def\vv{{\bf v}} 
\def\vA{{\bf A}} 
\def\vE{{\bf E}} 
\def\vF{{\bf F}} 
\def\vH{{\bf H}} 
\def\vJ{{\bf J}} 
\setcitestyle{numbers}
\begin{document}
\title{The Frequency Shift and Q of Disordered Superconducting RF Cavities}
\author{Hikaru Ueki}
\email{uekih@lsu.edu}
\author{Mehdi Zarea}
\email{zarea.mehdi@gmail.com}
\author{J. A. Sauls}
\email{sauls@lsu.edu}
\affiliation{Hearne Institute of Theoretical Physics, Louisiana State University,
             Baton Rouge, LA 70803}
\date{\today}
%-----------------------------------------------------------------------
\begin{abstract}
Niobium superconducting radio-frequency (SRF) cavities for high energy accelerator applications have been greatly improved in terms of the quality factor $Q$ by techniques such as Nitrogen doping. However, the mechanisms leading improvement in $Q$ are still not fully understood. Quite recently the SRF group at Fermilab measured anomalies in the frequency shift of N-doped SRF Niobium cavities near the transition temperature. Here we report our theoretical analysis of these results based on the microscopic theory of superconductivity that incorporates anisotropy of the superconducting gap and inhomogeneous disorder in the screening region of the SRF cavities. We are able to account for frequency shift anomalies very close to $T_c$ of the order of fractions of a kHz. Our results for the frequency shift and Q are in good agreement with the experimental data reported for all four N-doped Nb SRF cavities by Bafia et al.~\cite{baf21}.
We also compare our theory with an earlier report of on a Nb sample measured at 60 GHz~\cite{kle94a}.
In addition we show that the quality factor calculated theoretically has a peak of upper convexity with the largest $Q$ at intermediate levels of disorder. For strong disorder, i.e. the dirty limit, pair breaking in the presence of disorder and screening currents limits the $Q$. 
\end{abstract}
%------------------------------------------------------------------
\maketitle
%------------------------------------------------------------------
\section{Introduction}\label{sec-Introduction}

The performance of Niobium-based superconducting radio-frequency (SRF) cavities, as measured by the quality factor $Q$ and superheating field, $H_s$, for high energy physics accelerator technology have been improved significantly by techniques such as Nitrogen infusion~\cite{gra17}.
High-$Q$ SRF cavities are also being proposed and developed as detectors in the search for dark matter candidates. In recent proposals the sensitivity of the SRF cavity to axion-like dark matter detection depends on the square of the quality factor~\cite{bog19,gao21}.
SRF cavities operating at ultra-low temperatures and minimal microwave power have also attracted attention for applications in quantum information technology~\cite{rom20}.

However, doping and impurity infusion mechanisms leading to improved performance are not fully understood.
The roles of disorder and inhomogeneity on the current response near the vacuum-superconducting interface can lead to both subtle and competing 
effects~\cite{gur17,nga19}.
Given the range of applications of high-$Q$ SRF cavities a more detailed understanding of the effects of disorder on SRF performance is called for that combines first-principles theory of superconductivity, materials analysis and characterization.
Recent measurements of the resonant frequency of Nb-based SRF cavities with different surface treatments reveal non-monotonic temperature dependence (``anomalies'') in the shift of the resonant frequency of the cavity just below the superconducting transition, $\delta f=f_s(T)-f_n(T_c)$, for $|T-T_c|\ll T_c$. These anomalies are sensitive to surface treatment~\cite{baf21}. For Nitrogen-doped Nb SRF cavities these authors observed a negative slope in the resonance frequency just below the transition temperature, a minimum negative shift and rapid increase over a narrow temperature range just below $T_c$~\cite{baf21}.
Anomalies in the frequency shift just below $T_c$ for Nb coupled to a tunnel-diode oscillator operating at 10 MHz were reported much earlier~\cite{var74,var75}. 
Similar analysis of the impedance data of a Niobium sample embedded in a cylindrical copper cavity~\cite{kle94a} implies an anomaly in the resonance frequency shift at a nominal frequency of 60 GHz~\cite{baf19}.
The results for the frequency shift and $Q$ reported by Bafia et al. for N-doped Nb SRF cavities cover the frequency range $0.65-3.9\,\mbox{GHz}$~\cite{baf21}.

We report results based on the microscopic theory of nonequilibrium superconductivity~\cite{lar69,eli72,rai94}, combined with Slater's method for solving Maxwell's equations for the EM field confined in a closed metallic cavity~\cite{sla46}, for the frequency shift, $\delta f$, and quality factor $Q$ of the cavity as functions of temperature, frequency and disorder.
We are able to calculate frequency shifts of order $\delta f\sim 1-10^3\,\mbox{Hz}$ that exhibit non-monotonic behavior over a narrow temperature range, $|\Delta T|/T_c \lesssim 10^{-3}$, that are sensitive to impurity disorder. Indeed analysis of the frequency shift data near $T_c$ can provide a powerful method for measuring the scattering rate of electronic excitations by the random disorder potential.
In addition, the dependence of $Q$ as a function of the scattering rate from the random potential is calculated and shown to exhibit a peak of upper convexity with a maximum $Q$ obtained for cavities with intermediate levels of disorder, $1/2\pi\tau T_c\sim 1$. For strong disorder, $1/2\pi\tau T_c\gg 1$, pair breaking strongly suppresses $Q$~\footnote{We set $\kb=\hbar=1$, in which case
$1/\tau$, $T_c$ and $\omega$ and $\Delta$ have the same units. To covert to Gaussian units: $1/\tau T_c\rightarrow \hbar/\tau \kb T_c$, or $\omega\rightarrow\hbar\omega$, and use Gaussian units for $\kb$ and $\hbar$.} 
These results are discussed in Sec.~\ref{sec-Anomalies}.

Our theoretical predictions and analysis of $\delta f$ and $Q$ in the weak coupling limit for superconductors with disorder is in good agreement with experimental data reported for N-doped Niobium SRF cavities. An important element of the theory is the impact of disorder on the suppression of $T_c$ resulting from pair-breaking induced by the interplay of gap anisotropy and impurity scattering, i.e. weak violation of Anderson's theorem by non-magnetic disorder~\cite{zar22}. Inhomogeneous disorder then generates an inhomogeneous distribution for $T_c$ distributed throughout an SRF cavity. These results and comparision with experimental results on N-doped Nb cavities are reported in Sec.~\ref{sec_N-doped_Nb}.

We begin with the theoretical formulation and theoretical results for the a.c. conductivity of disordered superconductors, then connect the a.c. conductivity to the complex surface impedance in Sec.~\ref{sec-Current_Response}.
The key results of Slater's method connecting the surface impendance to the resonance frequency shift, $\delta f$ and $Q$ of the cavity are discussed in Sec.~\ref{sec-Surface_Impedance}.
The more technical aspects of the theoretical formalism are discussed in Appendix~\ref{appendix-Slater_Method}.
We use the formalism of Ref.~\onlinecite{rai94} for the a.c. conductivity in superconductors based on the quasiclassical theory of superconductivity. We incorporate gap anisotropy and inhomogeneous disorder to calculate the conductivity for superconducting RF cavities. We employ Slater's formulation for solving Maxwell's equations~\cite{gur17a,sla46} and derive relations between the conductivity and surface impedance in superconductors in the local limit, as well as expressions for the frequency shift and quality factor in Sec.~\ref{sec-Surface_Impedance}. We also derive the frequency shift near $T_c$ and show that the frequency shift has various types of anomalies near $T_c$ that are sensitive to the level of disorder. In Sec.~\ref{sec-Anomalies} we compare our calculations of the frequency shift and quality factor with the two experimental results of Klein et al.~\cite{kle94a,baf19} and Bafia et al.~\cite{baf21}. 

%-------------------------------------------------------------------
\begin{figure}\label{fig-EM_field_penetration}
\centering\includegraphics[width=0.85\columnwidth]{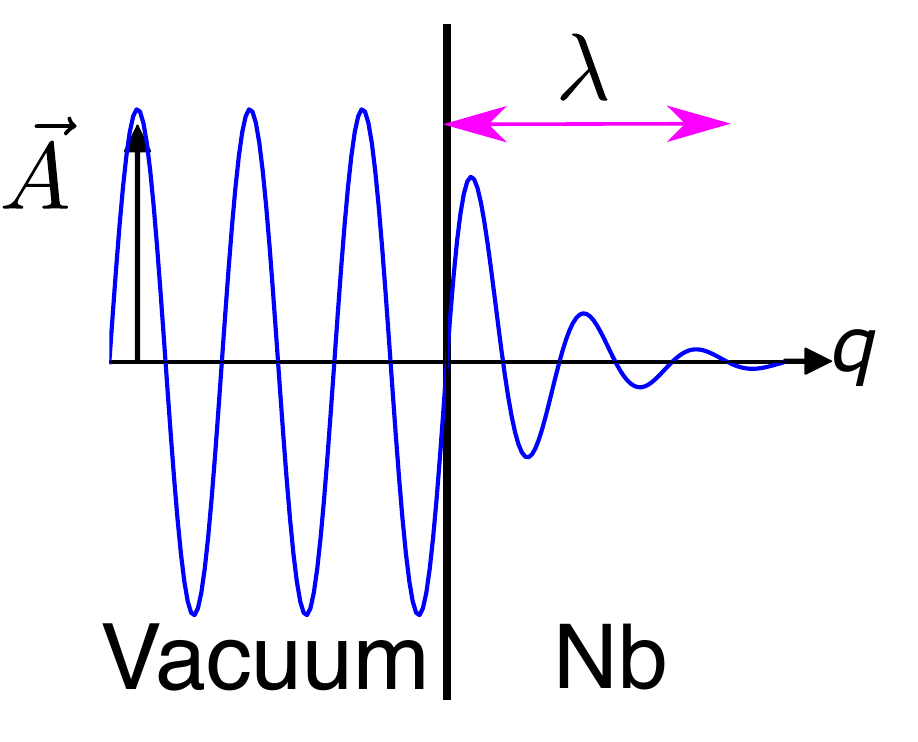}
\caption{
Transverse electromagnetic field penetration represented by the vector potential at a vacuum-superconductor interface. In the superconducting state field penetrations a distance of order the London penetration depth, $\lambda$, while in the normal state field penetration is governed by the skin depth, $\delta(\omega)=c/\sqrt{2\pi\omega\sigma_n}$, where $\sigma_n$ is the normal-state conductivity at frequency $\omega$.
}
\end{figure}
%-------------------------------------------------------------------

%----------------------------------------------------------
%\onecolumngrid
\begin{figure*}[t]
\includegraphics[width=\textwidth]{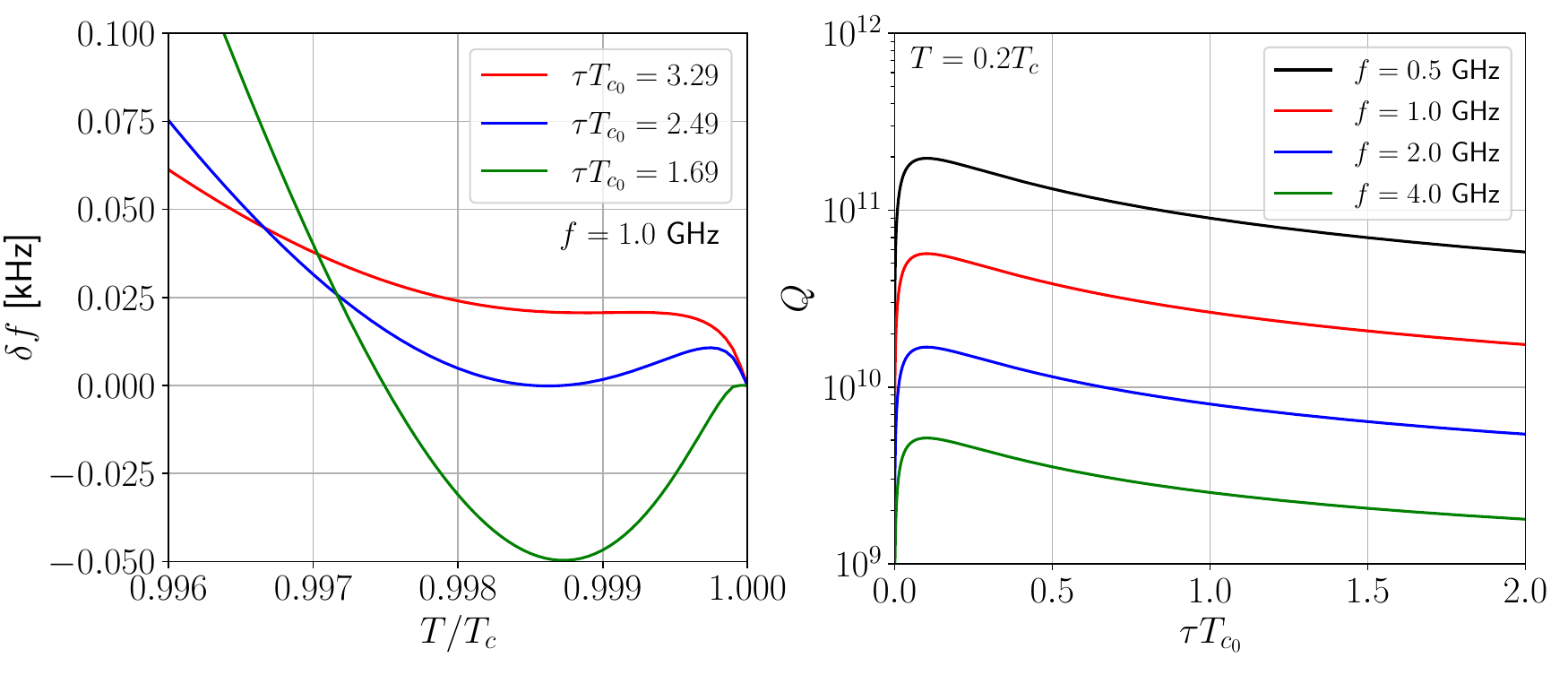}
\caption{
Left panel: Temperature dependence of the frequency shift $\delta f$ for relatively clean cavities with $f=1.0\,\mbox{GHz}$ near $T_c$.
Right panel: The effect of disorder on $Q$ at $T=0.2T_c$. 
Note the maximum $Q$ is near $\tau_0/\tau\approx 1.0$, i.e. 
intermediate disorder.
}
\label{fig-Delta_f_and_Q}
\end{figure*}
%\twocolumngrid
%----------------------------------------------------------

%-------------------------------------------------------------------
\section{Current response of an RF cavity}\label{sec-Current_Response}

Consider the current response and penetration of the electromagnetic field at the vacuum-metal interface of an SRF cavity as illustrated in Fig.~\ref{fig-EM_field_penetration}.
In this geometry the electric and magnetic fields are transverse to the interface normal, and the corresponding surface current for an isotropic superconductor takes the form, $\vJ(\vq,\omega)=K(\vq,\omega)\,\vA(\vq,\omega)$, where $K(\vq,\omega)$ is the linear response function determining the Fourier components of the current in terms of those of the EM field; $K(\vq,\omega)$ encodes the response of unbound quasiparticles, as well as the response of superconducting condensate, to the EM field and the random potential. 

Pure Niobium is a superconductor that is near the border between type I and type II magnetic behavior, i.e. with a London penetration depth, $\lambda$, comparable to the superconducting coherence length, $\xi$~\cite{pro22}. However, disorder increases the London penetration depth and also reduces the coherence length, driving Nb with disorder to type II behavior. In this limit the electromagnetic field penetrates relatively deep into the superconductor at a vacuum/superconducting interface and probes the {\it bulk} order parameter by 
exciting currents relatively far from the interface.
Thus, we negelect the surface deformation of the order parameter on the scale of $\xi$ and calculate the current response to the EM field in the local limit, in which case we can set $q=0$ in the response function, $K(\vq,\omega)$. The current response then reduces to 
\be\label{eq-current_response}
\vJ(\vq,\omega)=\sigma(\omega)\,\vE(\vq,\omega)
\,,
\ee
where $\sigma(\omega)\equiv-icK(0,\omega)/\omega$ is the a.c. conductivity.
To calculate $K(\vq,\omega)$, and thus $\sigma(\omega)$, we solve the non-equilibrium quasiclassical transport equations, originally formulated by Larkin and Ovchnnikov and Eliashberg~\cite{lar69,eli72}. Here we use the Keldysh formulation of nonequilibrium quasiclassical transport theory for strong-coupling and disordered superconductors in Ref.~\onlinecite{rai94}.
The current response is then given in terms of the of the $\tauz$-component the Keldysh propagator,
\begin{equation}\label{eq-Current-response}
\hspace*{-2mm}
\vJ(\vq,\omega)=N_f\int d\vp\,(e\vv_\vp)\int_{-\infty}^\infty\ns\ns d\varepsilon
\frac{1}{2}\Tr{\tz\whmfGk(\vp,\vq;\varepsilon,\omega)}
\,,
\end{equation}
where $\whmfGk(\vp,\vq;\varepsilon,\omega)$ is the solution of the Keldysh transport equations for the the Keldysh component of the nonequilibrium propogator. 
Note that $N_f$ is the total normal-state density of states at the Fermi energy, $e\vv_{\vp}$ is current carried by normal-state quasiparticles with Fermi velocity, $\vv_{\vp}$, for the point $\vp$ on the Fermi surface, and $e$ is the electron charge.
Pairing correlations and particle-hole coherence of the Bogoliubov excitations of the superconducting state are encoded in the Nambu structure of $\whmfGk$, the pairing self energy, $\whDelta$, and the Nambu matrix structure of the transport equations.

Here we consider the response of a conventional isotropic superconductor in the long-wavelength limit ($q\to0$) with disorder in the form of a random distribution of non-magnetic impurities. In the normal state the random potential leads to diffusive transport described by a mean transport scattering time, $\tau$, or mean free path, $\ell = v_f\tau$, where $v_f$ is the Fermi velocity for an isotropic metal.
For Nb the dimensionless phonon-mediated electron-electron interaction, $\lambda=1.057$, is spread over the phonon bandwidth of $\omega_{\mbox{\tiny D}}=27.3$ meV, and leads to $T_c=0.802$ meV, i.e. $T_c/\omega_{\mbox{\tiny D}}\simeq 0.029$, with corrections to the gap of order a few percent~\cite{zar22}. 
Thus for the purposes of calculating the current response we evaluate the equilibrium pairing self energy in the weak-coupling limit, $T_c\ll\omega_{\mbox{\tiny D}}$.
With these assumptions the current response function for a conventional superconductor given by Eq.~(68) of Ref.~[\onlinecite{rai94}] reduces to
\onecolumngrid
\begin{align}
\sigma (\omega) 
= 
&
\frac{\sigma_{\mbox{\tiny D}}}{i\omega\tau} 
\int_{-\infty}^\infty\frac{d\varepsilon}{4\pi i} 
\bigg\{
\tanh\left(\frac{\varepsilon-\omega/2}{2T}\right)
\frac{-\pi}{D^{\text{R}}(\varepsilon+\omega/2)+D^{\text{R}}(\varepsilon-\omega/2)+1/\tau}
\left[\frac{\varepsilon^2-\omega^2/4+\Delta^2}
{D^{\text{R}}(\varepsilon+\omega/2)D^{\text{R}}(\varepsilon-\omega/2)}
+1\right] 
\notag
\\
&
\hspace*{19mm}
-
\tanh\left(\frac{\varepsilon+\omega/2}{2T}\right)
\frac{-\pi}{D^{\text{A}}(\varepsilon+\omega/2)+D^{\text{A}}(\varepsilon-\omega/2)+1/\tau}
\left[\frac{\varepsilon^2-\omega^2/4+\Delta^2}
{D^{\text{A}}(\varepsilon+\omega/2)D^{\text{A}}(\varepsilon-\omega/2)}
+1\right] 
\label{eq-sigma}
\\
+\bigg[
&
\tanh\left(\frac{\varepsilon+\omega/2}{2T}\right)
-\tanh\left(\frac{\varepsilon-\omega/2}{2T}\right)\bigg] 
\times\frac{-\pi}{D^{\text{R}}(\varepsilon+\omega/2)+D^{\text{A}}(\varepsilon-\omega/2)+1/\tau}
\left[\frac{\varepsilon^2-\omega^2/4+\Delta^2}
{D^{\text{R}}(\varepsilon+\omega/2)D^{\text{A}}(\varepsilon-\omega/2)}
+1\right] 
\bigg\},
\notag
\end{align}
\twocolumngrid
\noindent where $\sigma_{\mbox{\tiny D}}\equiv(2/3)N_f e^2 v_f^2\tau$ is the Drude result for the d.c. conductivity of the normal metallic state, and $D^{\text{R,A}}(\varepsilon)$ are the retarded (R) and advanced (A) functions defined by $D^{\text{R,A}}(\varepsilon)\equiv\sqrt{\Delta^2-(\varepsilon\pm i\eta)^2}$, where $\eta\rightarrow 0^+$ and the upper and lower signs correspond to the retarded (R) and advanced (A) functions, respectively, that determine the corresponding non-equilibrium propagators. Note that there is no ``Dynes parameter'' introduced into the theoretical formalism, and the impurity scattering rate drops out of the equilibrium propagators, and thus $D^{\text{R,A}}(\varepsilon)$, for isotropic superconductors with non-magnetic impurities (Anderson's theorem). 
As a result the impurity scattering rate $1/\tau$ appears only explicitly as shown in the kernel of the conductivity in Eq.~\eqref{eq-sigma}~\cite{rai94}.
The three lines of this equation correspond, respectively, to the retarded, advanced and anomalous contributions to the nonequilibrium Keldysh kernel for the conductivity.
It is straight-forward to show that the normal-state limit ($\Delta\rightarrow 0$) of Eq.~\eqref{eq-sigma} reduces to the frequency-dependent Drude conductivity,
\begin{equation} 
\label{eq-sigma_n}
\sigma_n 
=
\sigma_{\mbox{\tiny D}}\frac{1+i\omega\tau}{1+(\omega\tau)^2}
\equiv
\sigma_{n_1} + i \sigma_{n_2}
\,.
\end{equation}
The imaginary part of the a.c. conductivity determines the normal metal skin depth. For $\omega\tau\ll 1$, $\delta(\omega)=c/\sqrt{2\pi\omega\sigma_{\mbox{\tiny D}}}$. The penetration of the EM field in the normal state, and thus the frequency shift of the cavity \emph{relative} to the geometric resonant frequency for a perfect conducting cavity, is independent of temperature in the vicinity near $T_c$. This shift determines the reference frequency for changes in the resonant frequency below $T_c$.   
The penetration depth changes rapidly as a function of temperature at the onset of superconductivity, leading to rapid changes in the resonant frequency, including anomalies very close to $T_c$.
The full analysis of both field penetration and the change in the resonant frequency of the cavity is presented in the following sections.

%--------------------------------------------------------------
\vspace*{5mm}
\section{Surface Impedance of SRF cavities}
\label{sec-Surface_Impedance}

To compare with experimental reports it is convenient to introduce the complex surface impedance of the vacuum-superconducting interface, 
\begin{align} 
Z_s \equiv Z_0 \frac{E_\parallel (z=0)}{H_\parallel (z=0)}
\,, 
\end{align}
where $E_\parallel (z=0)$ and $H_\parallel (z=0)$ are the electric and magnetic field components in the plane of the interface and evaluated just below the metal surface. The prefactor, $Z_0\equiv 4\pi/c=376.7\,{\rm \Omega}$, is the so-called impedance of the vacuum.
Maxwell's equations, combined with Eq.~\eqref{eq-current_response}, yield the surface impedance in terms of the complex a.c. conductivity~\cite{gur17a},
\begin{align}\label{eq-Zs}
Z_s = Z_0\sqrt{\frac{\omega}{4\pi i\sigma(\omega)}}
\,.
\end{align}
Equation~\eqref{eq-Zs} is valid for the surface response of superconductors in the local limit $\ell\ll\xi\ll\lambda$, where $\ell$, $\xi$, and $\lambda$ are the mean free path, coherence length, and the London penetration depth, respectively.
The same relationship holds for the surface impedance of the normal metal and the Drude conductivity, $Z_n = Z_0\sqrt{\frac{\omega}{4\pi i\sigma_n(\omega)}}$.
The conductivity and surface impedance are complex functions of frequency, temperature and scattering rate. Expressing $Z_s = R_s - i X_s$ in terms of the surface resistance, $R_s$ and surface reactance, $X_s$, we can relate these quantities to the real and imagniary components of the conductivity, $\sigma = \sigma_1 + i \sigma_2$. 

The relations connecting $R_s$ and $X_s$ to $\sigma_{1,2}$ simplify in the low-frequency limit, $\omega\tau \ll 1$, which is applicable to the analysis presented below for Nb SRF cavities with intermediate to strong disorder. In this limit $X_n \simeq R_n = Z_0\sqrt{f/4\sigma_{\mbox{\tiny D}}}$, and
\begin{eqnarray} 
\frac{R_s}{R_n} 
&=&\frac{\sigma_{n_1}^{1/2}}{(\sigma_1^2+\sigma_2^2)^{1/4}} 
\notag\\
&\times&
\left[
\cos\left(\frac{1}{2}\arctan\frac{\sigma_2}{\sigma_1}\right) 
-
\sin\left(\frac{1}{2}\arctan\frac{\sigma_2}{\sigma_1}\right)
\right]
\,,
\label{eq-Rs}
\\
\frac{X_s}{R_n} 
&=&\frac{\sigma_{n_1}^{1/2}}{(\sigma_1^2+\sigma_2^2)^{1/4}}
\notag\\
&\times&
\left[
\cos \left( \frac{1}{2} \arctan \frac{\sigma_2}{\sigma_1} \right) 
+
\sin \left( \frac{1}{2} \arctan \frac{\sigma_2}{\sigma_1}  \right) 
\right]
\,.
\label{eq-Xs}
\end{eqnarray} 

%------------ Fig. Comparison with Nb at 60GHz ---------------------
\begin{figure*}[t]
\includegraphics[width=\textwidth]{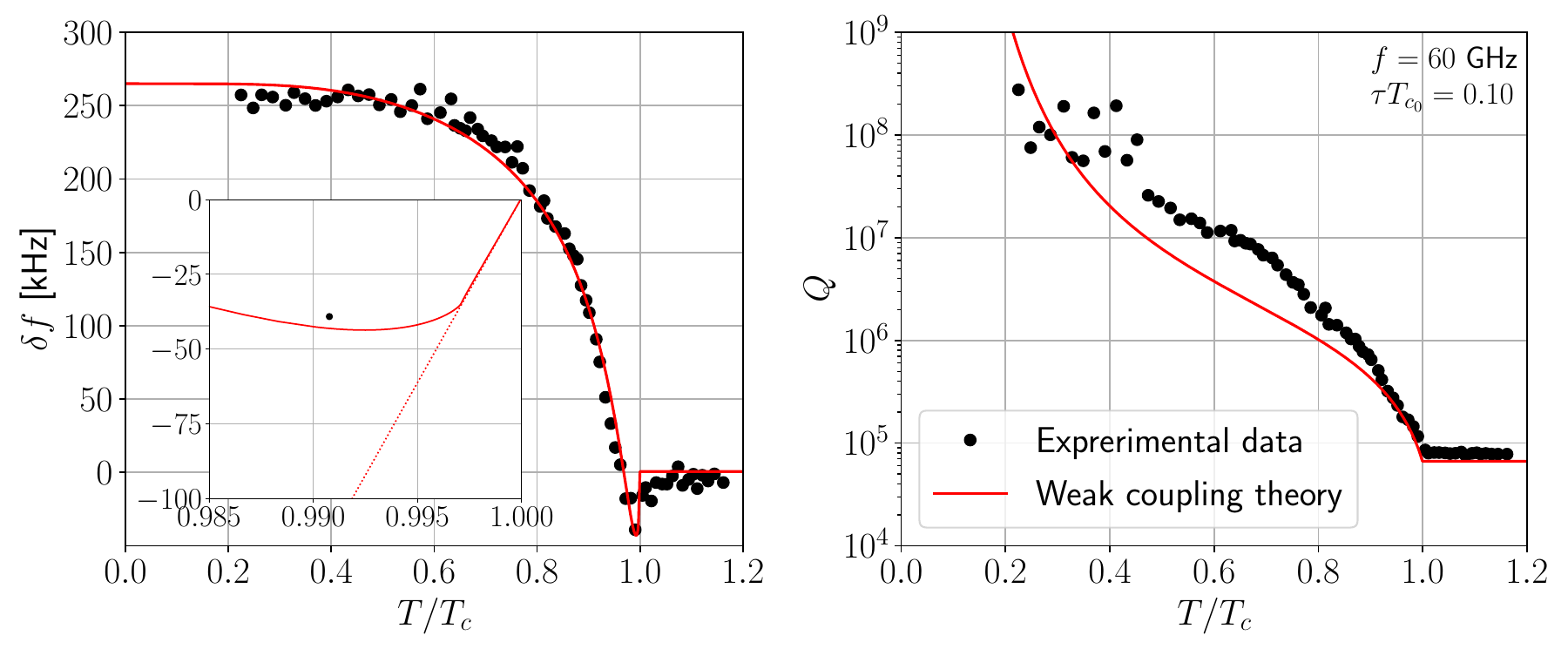}
\caption{Frequency shift $\delta f$ (left) and quality factor $Q$ (right) of a Nb sample embedded in a copper cavity as a function of temperature for a cavity with resonant frequency $f=60\,\mbox{GHz}$. The points are experimental data calculated from the surface impedance reported by Klein et al.~\cite{kle94a}.
The solid lines are calculated based on Eqs.~\eqref{eq-sigma},~\eqref{eq-Zs} and ~\eqref{eq-shift+Q_Slater} with the scattering rate $\tau_0/\tau =1.59$.
Inset: zoomed-in plot of $\delta f$ very near $T_c$. The dashed line is based on the perturbative expansion of the conductivity in $\Delta^2\propto (T_c-T)$ given by Eq.~\eqref{eq-Delta_f_near_Tc}.
}
\label{fig-Klein}
\end{figure*}
%-------------------------------------------------------------------

\subsection{$\delta f$ and $Q$ of SRF cavities}\label{sec-frequency_Q_SRF}

Slater solved Maxwell's equations in a hollow cavity with small damping and derived the expression for the quality factor and the frequency shift from an ideal cavity as~\cite{sla46} 
\begin{align} 
\frac{1}{Q}&+2i\frac{\Delta\omega_a}{\omega_a} =
i\frac{c}{\omega_a}\frac{\displaystyle\int_S(\vn\times\vE)\cdot\vH_ada}
{\displaystyle\int\vE\cdot\vE_adv} \notag \\
&-\frac{c}{\omega_a}\frac{\displaystyle\int_{S'}(\vn\times\vH)\cdot\vE_ada}
{\displaystyle\int\vE\cdot\vE_adv} 
+\frac{4\pi}{\omega_a}\frac{\displaystyle\int\vj\cdot\vE_adv}
{\displaystyle\int\vE\cdot\vE_adv}
\,,
\label{eq-Q_and_Delta_omega}
\end{align}
where $\omega_a$ is the angular frequency of the $a$th mode in an ideal resonant cavity, $\vE_a$ and $\vH_a$ are the $a$th mode of the electric and magnetic fields, respectively, $S$ and $S'$ denote conductive and insulating surfaces, respectively, and $\vn$ is the outer normal to the surface. See App. \ref{appendix-Slater_Method} for details. Here we consider only damping from metallic walls of an SRF cavity, and thus we neglect the second term on the right-hand side of Eq. (\ref{eq-Q_and_Delta_omega}). We also can neglect the third term since the source current vanishes in the vacuum region. We substitute $\vn\times\vE=(Z_s/Z_0)\vH$ into Eq.~\eqref{eq-Q_and_Delta_omega}, and assume that $\vH$ is given by the magnetic field of the $a$th mode~\cite{sla46}. Finally, we use $\int\vE\cdot\vE_adv=i\int\vH\cdot\vH_adv$ from App.~\ref{appendix-Slater_Method}.
\begin{align} 
\vH=\vH_a\int\vH\cdot\vH_a\,dv =-i\vH_a\int\vE\cdot\vE_a\, dv
\,. 
\end{align}
As a result we obtain the quality factor and frequency shift in terms of the surface impedance
\begin{align}
\label{eq-shift+Q_Slater} 
\frac{1}{Q}+2i\frac{\Delta\omega_a}{\omega_a}=\frac{Z_s}{G}, 
\end{align}
where $G$ is the geometric factor of the cavity defined by $G\equiv Z_0\omega_a/c\int_S\vH_a^2da$. 
Using the Eq.~\eqref{eq-shift+Q_Slater} Slater calculated the frequency shift and quality factor for a normal-metal-vacuum interface~\cite{sla46}. Furthermore, using $\omega_a=2\pi f_a$ and subtracting the normal frequency shift from the superconducting frequency shift, we obtain relation between the frequency shift of the superconducting cavity measured relative to the normal-state resonance frequency and the corresponding surface reactances,
\begin{equation} 
\delta f=\frac{f}{2G}(X_n-X_s)\,,\quad
G=Z_0\frac{2\pi f}{c}\frac{\displaystyle\int\vH^2dv}{\displaystyle\int_S\vH^2da}.
\label{eq-Delta_f}
\end{equation}

The subscript $a$ for the relevant mode is omitted hereafter. This formula is used to calcuate the frequency shift of a superconducting sample in a cavity. We also obtain the quality factor from the surface resistance,
\begin{equation} 
Q=\frac{G}{R_{\rm s}}.
\label{eq-Q}
\end{equation}

\subsection{Material parameters for Nb}

From the Drude conductivity the normal-state surface resistance can be related to the scattering rate, $1/\tau$, and the plasma frequency, $\omega_p$, by $\sigma_{\mbox{\tiny D}}=\frac{\omega_p^2\tau}{4\pi}$. Thus we have,
\begin{align} 
R_n=\frac{Z_0}{\omega_p}\sqrt{\frac{\pi f}{\tau}}
\,.
\label{eq-Rn}
\end{align}
The plasma frequency, $\omega_p=\sqrt{(8\pi/3) N_f e^2 v_f^2}$, is also simply related to the zero-temperature limit of the London penetration depth for pure Nb, $\lambda_{\mbox{\tiny L}}=c/\omega_p$, both expressed in terms of the electron charge, $e$, density of states $N_f$, Fermi velocity, $v_f$. 
We adopt the Fermi velocity reported in Ref.~\onlinecite{max65}, $v_f=0.26\times10^8\,\mbox{cm/s}$, and the London penetration depth from Ref.~\onlinecite{kle94a}. With these parameters we calculate the density of states, $N_f=0.101\,\mbox{eV}^{-1}\,{\AA}^{-3}$, and the plasma frequency, $\omega_p=c/\lambda_{\text{L}}=9.08\times10^6\,\mbox{GHz}$.
We use these results and the normal-state surface resistance as input to calculations of the frequency shift and quality factor of SRF cavities.

%----------------------------------------------------------------
\section{Anomalies in $\delta f$ and Maximum $Q$}
\label{sec-Anomalies}

We highlight the two key features of the cavity response for relatively clean SRF cavities with $f=1.0\,\mbox{GHz}$ and $G=250\,{\rm \Omega}$: (i) the non-monotonic behavior of the frequency shift and its sensitivity to the level of disorder, and (ii) the maximum in the quality factor at finite levels of disorder. Both features are shown in 
Fig.~\ref{fig-Delta_f_and_Q}. The left panel shows three general features in $\delta f$ for temperatures very near $T_c$ depending on the scattering rate $1/\tau$: (a) a negative shift, then a recovery (dip feature) for $\tau_0/\tau=9.42\times 10^{-2}$, (b) a positive shift, then a decline and recovery (bump feature) for $\tau_0/\tau =6.39\times 10^{-2}$, and finally (c) a rapid increase followed by a slow recovery (plateau feature) for $\tau_0/\tau =4.84\times 10^{-2}$, where $\tau_0=1/2\pi T_{c_0}$ is the pair formation timescale and $T_{c_0}=9.33\,\mbox{K}$ is the transition temperature of pure Nb~\cite{zar22}.

All three types of frequency shift anomalies are observed in Nb SRF cavities depending on the levels of disorder~\cite{baf21}.
We also find that the quality factor $Q$ has a peak of upper convexity as a function of the mean scattering time, confirming results by other authors~\cite{tur91,ben99,padamsee08}. Thus, the maximum $Q$ in SRF cavities occurs for intermediate levels of disorder, and is suppressed in dirty cavities due to the pair breaking.

%----------------------------------------------------------------
\subsection{Analysis of Nb in a Cooper Cavity at $f=60\,\mbox{GHz}$}

Figure~\ref{fig-Klein} shows the temperature dependence of the frequency shift and quality factor for a Nb sample embedded in a copper cavity. The points are experimental data calculated from the surface resistance and reactance reported by Klein et al.~\cite{kle94a}.
We have calculated the a.c. conductivity based on the weak coupling limit for the conductivity given by Eq.~\eqref{eq-sigma}. We then calculate quality factor and frequency shift obtained from the conductivity and compare with the experimental results for a Niobium sample in a copper cavity as reported and measured by Klein et al.~\cite{kle94a}. The resonance frequency and the resonator constant of the copper cavity used here were $f=60\,\mbox{GHz}$ and $\nu=Z_0/2G=0.0458$, respectively. 
The solid lines are the theoretical results. The monotonic increase in $Q$ is expected, although the comparison between theory and experiment is not perfect, but the experimental data does have significant scatter~\cite{kle94a}.
For the frequency shift the comparison is much better. In addition we obtain theoretically the negative frequency shift anomaly that is implicit in the experimental data for the reactance in the close vicinity of $T_c$.
The inset shown in the left panel of Fig.~\ref{fig-Klein} shows a zoom-in of the region very near $T_c$, with the exact result for the shift shown as the solid line, which is in agreement with the data from Klein et al.~\cite{kle94a}. The dashed line is the result obtained by perturbation theory in $\Delta^2\propto (T_c-T)$, which is valid only for $\omega > 2\Delta(T)$. In this region $\sigma_1^{(2)}$ defined by Eq.~\eqref{eq-sigma1_and_sigma2_near_Tc} is negative since pair breaking results from absorption in this window near $T_c$.

\noindent In addition, $\sigma_2^{(2)}-\sigma_1^{(2)}>0$ implying that the frequency shift is negative just below $T_c$. The authors of Ref.~\onlinecite{kle94a} compared their results with calculations of the conductivity based on Mattis--Bardeen theory in the dirty limit~\cite{mat58}. They adjusted $T_c$ and the gap to fit the experimental data. 
Our analysis incorporates the roles of gap anisotropy and disorder to account for both the suppression of $T_c$ and the gap amplitude in order to obtain a consistent theory for the effects of disorder on $T_c$, $\Delta$ and the conductivity as a function of the quasiparticle scattering rate $1/\tau$. Our analysis yields 
$\tau_0/\tau \simeq 1.6$, implying that the samples studied were in the cross-over region between weak and strong disorder~\footnote{The dimensionaless scattering rate can be expressed in terms of other known or measured parameters: $2\pi\tau T_{c_0}\rightarrow2\pi\tau \kb T_{c_0}/\hbar=(\ell/\xi_0)$,
where $\ell=v_f\tau$ is the elastic mean free path in the normal metal, 
and $\xi_0=\hbar v_f/2\pi\kb T_{c_0}$ is the the superconducting coherence length in the clean limit.}.

%----------------------------------------------------------
\begin{figure*}[t]
\includegraphics[width=\textwidth]{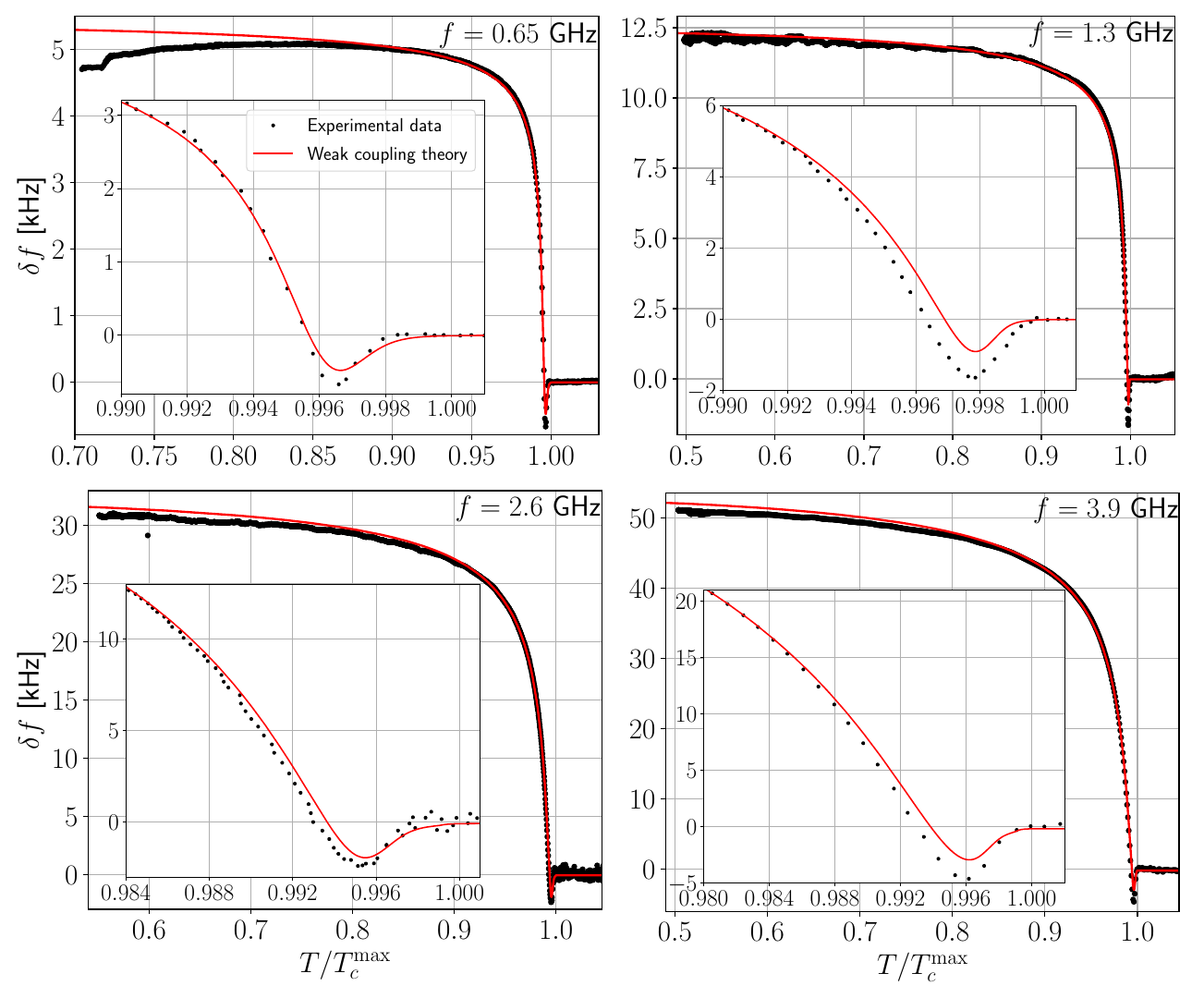}
\caption{Temperature dependence of the frequency shift $\delta f$ of the Nb SRF cavities with each different resonance frequency, $f=0.65\,\mbox{GHz}$ (upper left), $1.3\,\mbox{GHz}$ (upper right), $2.6\,\mbox{GHz}$ (lower left), and $3.9\,\mbox{GHz}$ (lower right), near $T_c$. The black points are the measured values extracted from and reported by Bafia et al.~\cite{baf21}, and the red lines are our calculations based on weak coupling theory with inhomogeneous disorder. Each inset is the zoomed-in region near $T_c$.}
\label{fig-Bafia}
\end{figure*}
%----------------------------------------------------------

%----------------------------------------------------------------
\subsection{Effects of anisotropy and disorder on $T_c$, $\Delta$ and $\sigma(\omega,T)$}

Niobium is an anisotropic superconductor belonging to the identity representation of the cubic point group. Thus, the gap amplitude varies with position (Fermi momentum) on the Fermi surface, but is otherwise invariant under the point group symmetry. 
The combination of anisotropy of the Cooper pair amplitude defined on the Fermi surface in momentum space and elastic scattering by non-magnetic disorder, from either impurities or structural defects, leads to the violation of Anderson's theorem, i.e. the suppression of pair formation and reduction of $T_c$, as well as the maximum excitation gap. The suppression of $T_c$ has been calculated and is given by~\cite{zar22}
\begin{equation}\label{eq-Tc} 
\ln\frac{T_{c_0}}{T_c}
=
\cA\times 2\pi T_c
\sum_{\varepsilon_n>0}^\infty
\left(
\frac{
\hbar/\tau
}
{\varepsilon_n(\varepsilon_n+\nicefrac{1}{2}\hbar/\tau)
}
\right)
\,,
\end{equation}
where $\varepsilon_n=(2n+1)\pi T_c$ are the Matsubara energies evaluated at the transition temperature for disordered Nb. 
The gap amplitude has the general form, $\Delta(\vp)=\Delta(T)\cY(\vp)$, where $\cY(\vp)$ is Cooper pair amplitude as a function of position $\vp$ on the Fermi surface, normalized to $\int d\vp\,|\cY(\vp)|^2 = 1$, The integration is an average over the Fermi suface. In the limit $T\rightarrow T_c$ the Cooper pair amplitude, $\cY(\vp)$, is the eigenfunction of the linearized gap equation that determines the $T_{c_0}$ in the clean limit.
The anisotropy parameter in Eq.~\eqref{eq-Tc} is defined by the limit 
\begin{equation}
\cA= \lim_{T\rightarrow T_{c_0}}
\frac{\langle|\Delta(\vp)|^2\rangle-|\langle\Delta(\vp)\rangle|^2} 
     {\langle|\Delta(\vp)|^2\rangle}
=
1-|\langle\cY(\vp)\rangle|^2, 
\end{equation}
where $\langle\cdots\rangle = \int d\vp\cdots$ is the the Fermi surface average with $\langle 1 \rangle= 1$ and $\cY(\vp)$ is normalized as $\langle|\cY(\vp)|^2\rangle=1$. In general the anisotropy factor $\cA$ can range from $\cA=0$ for an isotropic $s$-wave superconductor, to the other extreme of $\cA=1$, i.e. $\langle\Delta(\vp)\rangle\equiv 0$, characteristic of unconventional superconductors that break the point group symmetry. For conventional, anisotropic superconductors $0 < \cA < 1$, and for Niobium we use the results of a recent first-principles calculation of the gap anisotropy parameter for pure single crystalline Niobium of $\cA=0.037$ and $T_{c_0}=9.33\,\mbox{K}$~\cite{zar22}. Figure~11 in Ref.~\cite{zar22} shows the suppression of $T_c$ for Nb as a function of the dimensionless scattering rate $1/2\pi\tau T_{c_0}$.

While the role of anisotropy is essential in accounting for the variations of $T_c$ and the excitation gap with disorder, the a.c. conductivity, at least for fully gapped conventional superconductors, is otherwise insensitive to the gap anistropy for frequencies, $\hbar\omega\ll 2\Delta(\vp)$.
Thus, we neglected the explicit dependence on gap anisotropy in Eq.~\eqref{eq-sigma} for the conductivity, but we include the implicit dependence on anisotropy in the suppression of $T_c$ and $\Delta(T)\equiv\sqrt{\langle|\Delta(\vp)|^2\rangle}$. We can calculate $\Delta(T)$ from the BCS gap equation, expressed below in terms of the zero temperature gap, $\Delta_0=\pi{\rm e}^{-\gamma}T_c$ where $\gamma=0.57721\cdots$ is the Euler-Mascheroni constant,
\begin{align} 
\ln \frac{\Delta}{\Delta_0} 
= 2\int_0^\infty d\xi\frac{1}{\sqrt{\xi^2+\Delta^2}}
\frac{1}{e^{\sqrt{\xi^2+\Delta^2}/2T}+1}
\,.
\label{eq-gap}
\end{align}
The solution for the BCS gap function is well described by the interpolation formula
\begin{align} 
\Delta(T,T_c)=\Delta_0\tanh\left(\frac{\pi T_c}{\Delta_0}
\sqrt{\frac{2}{3}\frac{\Delta C}{C_n(T_c)}\frac{T_c-T}{T}}\right)
\Theta(T_c-T), 
\label{eq-interpolation}
\end{align}
where $\Delta C\equiv C(T_c)-C_n(T_c)=12\,C_n(T_c)/7\zeta(3)$ is the jump in the heat capacity at $T_c$, with $C(T_c)$ and $C_n(T_c)$ denoting the superconducting and normal-state heat capacities, respectively, and $\zeta(3)=1.20205\cdots$ is the Riemann zeta function. We use Eq.~\eqref{eq-interpolation} in the analysis to follow to incorporate the effects of inhomogeneous disorder on $T_c$ and $\Delta(T)$.

%----------------------------------------------------------
\begin{figure*}[t]
\includegraphics[width=0.9\linewidth]{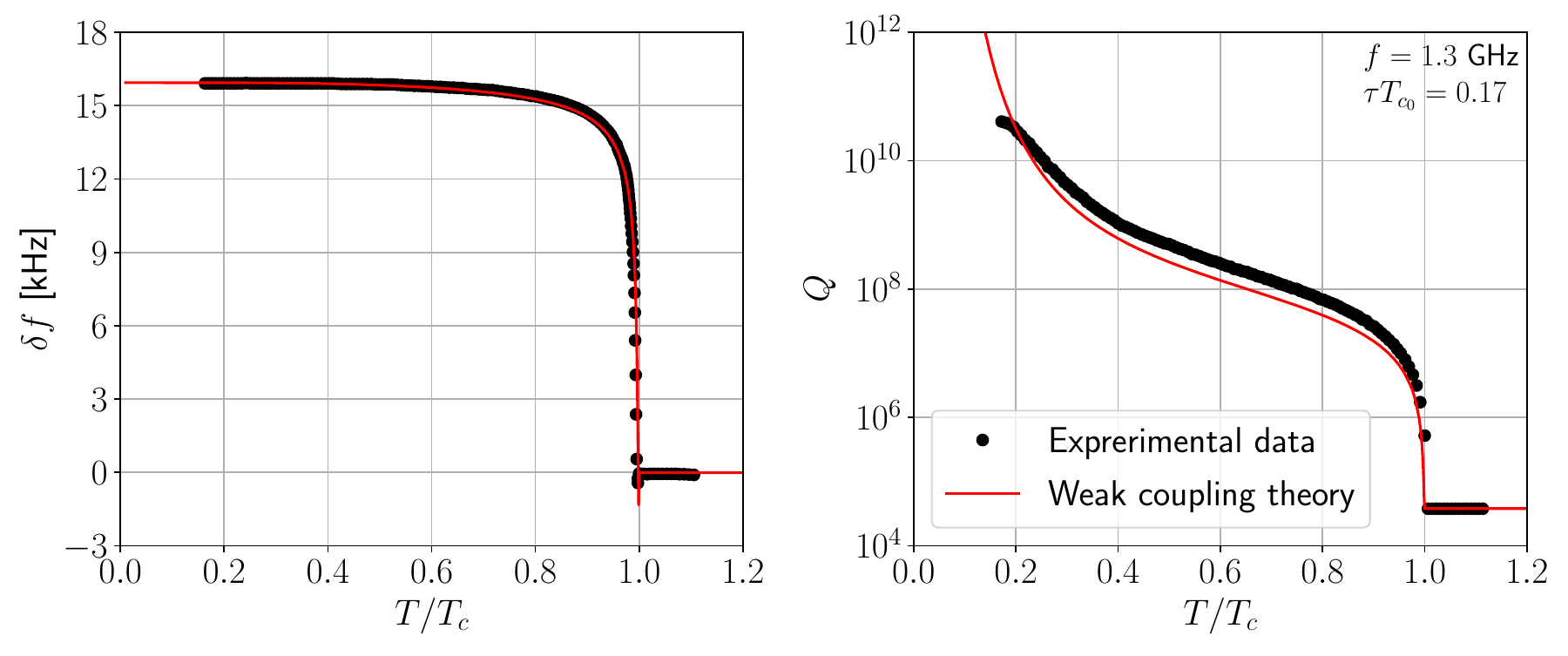}
\caption{Frequency shift $\delta f$ (left) and quality factor $Q$ (right) of an SRF cavity with nominal resonant frequency $f_0=1.3\,\mbox{GHz}$ as a function of temperature. The points are experimental data calculated from the surface impedance reported by Bafia et al.~\cite{baf21}, and the lines are theoretical results calculated with mean scattering rate  
%$\tau T_{c_0}=0.17$ 
$\tau_0/\tau=0.936$ 
based on weak coupling theory.}
\label{fig-Bafia-Q}
\end{figure*}
%----------------------------------------------------------

%----------------------------------------------------------------
\subsection{Analysis of N-doped Nb SRF Cavities}
\label{sec_N-doped_Nb}

Detailed measurements of the temperature dependence of the frequency shift $\delta f$ are reported for four Nitrogen-doped Nb SRF cavities with characteristic frequencies of $f_0 = \{0.65,1.3,2.6,3.9\}\,\mbox{GHz}$, including the the region of temperature very near $T_c$~\cite{baf21}. 
The experimental data, taken from Ref.~\onlinecite{baf21} and reproduced in Fig.~\ref{fig-Bafia}, shows a negative frequency shift anomaly in a narrow region of tempreature that depends of $f_0$, $T_c$ and $1/\tau$, as well as inhomogeneity of the disorder and its effect on $T_c$ and $\Delta(T,T_c)$~\footnote{Cavities with other surface treatments also show frequency shift anomalies near $T_c$, albeit over much wider temperature ranges [c.f. Ref.~\cite{ito21}].}
Table~\ref{tbl-Bafia} summarizes the key parameters of the four N-doped Nb cavities, including the spread in $T_c$ measured locally for each cavity~\cite{bafpc}.

We calculated the conductivity from Eq.~\eqref{eq-sigma}, adjusting the scattering rate $1/\tau$ to obtain an optimal fit for $\delta f(T)$ for each cavity.
The geometric factors for each of the Tesla cavities were provided by the Fermilab team:
$G=255 \ {\rm \Omega}$ for $f=0.65\,\mbox{GHz}$, 
$G=270 \ {\rm \Omega}$ for $f=1.3\,\mbox{GHz}$, 
$G=270 \ {\rm \Omega}$ for $f=2.6\,\mbox{GHz}$, 
and 
$G=293 \ {\rm \Omega}$ for $f=3.9\,\mbox{GHz}$~\cite{bafpc}.
From this input and our theoretical formulation we calculate the a.c. conductivity and obtain the negative frequency shift anomaly of the right magnitude. However, a single parameter description of the disorder, i.e. {\it homogeneous disorder}, does not accurately predict the temperature at the minimum of $\delta f$.
From the spread in $T_c$ for each cavity we construct a distribution profile, $\rho(T_c)$ representing the inhomogeneity of the disorder for each cavity. 
Introducing $T_c^{\text{max}}$ and $T_c^{\text{min}}$, we express the average and variance in the Gaussian distribution of $T_c$ as $\mu=T_c^{\text{av}}\equiv(T_c^{\text{max}}+T_c^{\text{min}})/2$ and $\mu_2=[(T_c^{\text{ave}}-T_c^{\text{min}})/3]^2$, respectively. We determine $T_c^{\text{max}}$ and $T_c^{\text{min}}$ as the two key fitting parameters so that the calculation of the frequency shift best fits the experimental data. The corresponding distribution for the scattering rate can be obtained using Eq.~\eqref{eq-Tc}. The details of this analysis for each cavity is described in App.~\ref{appendix-Inhomogeneous_Disorder} with the corresponding plots of the distributions of $T_c$, $1/\tau$ as well as the inhomogeneously broadened gap function defined by
\begin{equation} 
\Delta(T)=\Delta_0\sqrt{\int_{0}^{+\infty}\,dT_c\,\rho(T_c)\Delta^2(T,T_c)}
\,.
\label{eq-gap_spread_in_Tc}
\end{equation}
Note that overall magnitude of the disorder as well as the spread varies among the cavities, although all exihibit a similar level of disorder with typical scattering rates ranging between $\sim 4-6\,\mbox{ps}^{-1}$. Note also that the inhomogeneously broadened gap function rises more slowly in temperature due to the spread in $T_c$, leading to a change in the position of the minimum in the frequency shift anomaly.
We use the average value of $1/\tau$ obtained from $\tilde\rho(1/\tau)$ shown in Fig.~\ref{fig-inhomogeneity} in Eqs.\eqref{eq-sigma} and \eqref{eq-Rn} in order to calculate the conductivity and normal-state surface resistance. In particular, 
%$\bar\tau T_{c_0}=
$\bar\tau/\tau_0=
%0.214$, 
1.345$,
%$0.273$, 
$1.715$,
%$0.300$, 
$1.885$,
and 
%$0.299$ 
$1.879$
for $f=0.65\,\mbox{GHz}$, $1.3\,\mbox{GHz}$, $2.6\,\mbox{GHz}$, and $3.9\,\mbox{GHz}$, respectively, where $1/\bar\tau$ is the average of $1/\tau$. Note that we can also obtain the average of $1/\tau$ by solving Eq.~\eqref{eq-Tc} at the average value of $T_c$. 
Table \ref{tbl-Bafia} shows $T_c^{\text{max}}$, $T_c^{\text{min}}$, and $R_n$, used for the calculation based on the weak coupling theory with the spread in $T_c$, as well as the measured values from Ref.\onlinecite{bafpc}, for the four N-doped Nb SRF cavities.\footnote{According to Ref.~\onlinecite{bafpc} the temperature was measured at the equator, the upper beam pipe, and the lower beam pipe of each cavity using several Cernox RTD temperature sensors, with an observed spread in $T_c$ for each cavity~\cite{bafpc}.} As shown in Table~\ref{tbl-Bafia}, the experimental and theoretical values for $T_c^{\text{max}}$, $T_c^{\text{min}}$, and $R_n$ are in good agreement with each other.

%----------------------------------------------------------
\begin{table}[t]
\begin{center}
\caption{$T_c^{\rm max}$, $T_c^{\rm min}$, and $R_n$ 
used for the calculation based on the weak coupling theory 
with the spread in $T_c$ (second row), 
and measured by Bafia et al. (third row), for the Nb SRF cavities 
with each different resonance frequency.}
\begin{ruledtabular}
\begin{tabular}{c|cccc|cccc}  
 &\multicolumn{4}{c|}{Theory} & \multicolumn{4}{c}{Experiment} \\ \hline
$f \ [\mbox{GHz}]$ & 0.65 & 1.3 & 2.6 & 3.9 & 0.65 & 1.3 & 2.6 & 3.9 \\ \hline 
$T_c^{\rm max} \ [\mbox{K}]$ & 8.965 & 9.004 & 9.044 & 9.032 & 9.005 & 8.907 & 9.081 & 9.165 \\
$T_c^{\rm min} \ [\mbox{K}]$ & 8.895 & 8.976 & 8.980 & 8.990 & 8.975 & 8.87 & 9.041 & 9.15 \\
$R_n \ [{\rm m\Omega}]$ & 4.471 & 5.601 & 7.554 & 9.272 & 4.364 & 5.425 & 6.95 & 8.93 \\
\end{tabular}
\end{ruledtabular}
\label{tbl-Bafia}
\end{center}
\end{table}
%----------------------------------------------------------

Figure \ref{fig-Bafia} shows the temperature dependence of the frequency shift of the four N-doped Nb SRF cavities measured by Bafia et al.~\cite{baf21}, and the comparison with our calculated result based on Eq.~\eqref{eq-Delta_f} and weak-coupling theory for the conductivity, including inhomogeneous disorder with the spread in $T_c$ for the N-doped Nb SRF cavities.
The origin of the negative frequency shift anomaly onsetting at $T_c$ is the same as described earlier in the context of the analysis of the results of Klein et al.~\cite{kle94a}.
The negative shift and `dip feature' in the frequency shift follows from the change in the surface reactance defined in Eq.\eqref{eq-Xs}; $X_s/R_n$, which is proportional to the frequency shift with the negative proportionality coefficient, is the product of $[(\sigma_1/\sigma_{1n})^2+(\sigma_2/\sigma_{1n})^2]^{-1/4}$ and $\cos \left[ \frac{1}{2} \arctan (\sigma_2/\sigma_1) \right] + \sin \left[ \frac{1}{2} \arctan (\sigma_2/\sigma_1) \right]$ as seen in Eq. (\ref{eq-Xs}), and the former decreases and the latter increases as temperature is lowered below $T_c$.  Therefore, $X_{\rm s}/R_{\rm n}$ and the frequency shift have a peak near $T_c$ when $\sigma_2/\sigma_1$ in the latter increases sharply below $T_c$.

We also see from Eq.~\eqref{eq-Rs} that the ratio $\sigma_2/\sigma_1$ is important in determining the temperature and disorder dependence of the quality factor. In particular, $R_s/R_n$, which is inversely proportional to the quality factor, is the product of $[(\sigma_1/\sigma_{1n})^2+(\sigma_2/\sigma_{1n})^2]^{-1/4}$ and $\cos \left[\frac{1}{2} \arctan (\sigma_2/\sigma_1) \right] - \sin \left[ \frac{1}{2} \arctan(\sigma_2/\sigma_1) \right]$, both the former and latter decrease as the temperature drops below $T_c$, and the latter is the dominant contribution to $R_s/R_n$. Thus, since $R_s/R_n$ decreases more drastically when $\sigma_2/\sigma_1$ increases more sharply below $T_c$, the quality factor increases. 

We also note that the drop in $\delta f$ at the lower temperatures shown for $f=0.65$ GHz that deviates from the theoretical curve may reflect the presence of subgap states near the Fermi level, that are not present in the theory presented, that lead to a non-monotonic increase in the penetration depth, and thus a decrease in resonance frequency.

In summary, both the quality factor and the negative frequency shift near $T_c$ are enhanced by the large value of $\sigma_2$ for intermediate levels of disorder. 
To check if our theory also predicts the experimentally reported quality factor $Q(T)$, we compare our theory with the data reported in Fig.~2 of the Supplemental Material of Ref.~\onlinecite{baf21}. Figure \ref{fig-Bafia-Q} shows the comparison for the temperature dependence of the frequency shift and quality factor of a different N-doped Nb SRF cavity with $f=1.3\,\mbox{GHz}$.
The data points were obtained from the surface impedance reported in the Supplemental Material for Ref.~\onlinecite{baf21}, and the solid lines are the calculated results 
with 
%$\tau T_{c_0}=0.17$ 
$\tau_0/\tau=0.936$ 
based on weak coupling theory and homogeneous disorder, i.e. neglecting the spread in $1/\tau$ and $T_c$. The frequency shift is in close agreement, while the quality factor obtained by our theory is not perfect, but in reasonable agreement with that of Ref.~\onlinecite{baf21}. 
If the low-temperature behavior shown in the experimental result for $Q(T)$ is reproducible then there is likely low-energy sub-gap states present in this cavity that are not present in the theory we present.

\section{Conclusions} \label{sec-Conclusions}

We developed a theory and computational method to calculate the frequency shift $\delta f$, quality factor $Q$, and surface impedance $Z_s$, from the a.c. conductivity $\sigma$ of disordered superconducitng RF cavities based on the quasiclassical theory of superconductivity in the weak coupling limit, combined with Maxwell's equations. We showed that the observable properties of these cavities are well describe by the weak coupling theory with an anisotropic gap function and scattering by a non-magnetic distribution of disorder. We show that $\delta f$ vs $T\lesssim T_c$ and $T_c$ vs $1/\tau$ provide sensitive measures of the quality of disordered Nb cavities. Our analysis also indicates that even ``state of the art'' N-doped Nb SRF cavities can be further improved in terms of $Q$ by reducing the overall level of disorder. However, there is a level of disorder that is desirable because the quality factor has a maximum as a function of the quasiparticle-impurity scattering rate, with the largest $Q$ corresponding to intermediate disorder. For strong disorder, i.e. the dirty limit, pair breaking limits $Q$. 

\section{Acknowledgements}

We thank Daniel Bafia, Anna Grassellino, Alex Romanenko and John Zasadzinski for many discussions on their experimental work on SRF cavities, and for motivating this study. 
The research of HU and MZ was supported by National Science Foundation Grant PHY-1734332 and the Northwestern-Fermilab Center for Applied Physics \& Superconducting Technologies. The work of JAS is supported by the U.S. Department of Energy, Office of Science, National Quantum Information Science Research Centers, Superconducting Quantum Materials and Systems Center (SQMS) under contract number DE-AC02-07CH11359.

\appendix

\section{Slater's Method}\label{appendix-Slater_Method}

We consider the electromagnetic field in a hollow cavity to derive Eq.~\eqref{eq-Q_and_Delta_omega} obtained by Slater~\cite{sla46}. We first expand the EM fields as 
\begin{subequations} 
\begin{align} 
\vE&=\sum_a\left(\vE_a\int\vE\cdot\vE_adv
+\vF_a\int\vE\cdot\vF_adv\right), \\
\vH&=\sum_a\vH_a\int\vH\cdot\vH_adv, \\
\vj&=\sum_a\left(\vE_a\int\vj\cdot\vE_adv
+\vF_a \int\vj\cdot\vF_adv\right), \\
\rho&=\sum_a\psi_a\int\rho\psi_adv,
\end{align}\label{eq-expansion1}
\end{subequations} 
using the solenoidal vector functions $\vE_a$ and $\vH_a$, which satisfy the following equations: $k_a\vE_a=\nabla\times\vH_a$ and $k_a\vH_a=\nabla\times\vE_a$, and the irrotational vector function $\vF_a$, which is the gradient of a scalar function $\psi_a$, $k_a\vF_a=\nabla\psi_a$, where $k_a$ is the propagation constant associated with the $a$th mode as discussed below. These functions satisfy the normalization and orthogonality conditions as 
\begin{subequations} 
\begin{align} 
\int\vE_a\cdot\vE_bdv&=\delta_{ab}, \ \ \ 
\int\vH_a\cdot\vH_bdv=\delta_{ab}, \\
\int\vF_a\cdot\vF_bdv&=\delta_{ab}, \ \ \ 
\int\psi_a\psi_bdv=\delta_{ab}, \\
\int\vE_a\cdot\vF_bdv&=0, 
\end{align}
\end{subequations} 
wave equations,
\begin{subequations} 
\begin{align} 
\nabla^2\vE_a+k_a^2&\vE_a={\bf 0}, \ \ \
\nabla^2\vH_a+k_a^2\vH_a={\bf 0},  \\
&\nabla^2\psi_a+k_a^2\psi_a=0, 
\end{align}\label{eq-wave_eq}%
\end{subequations} 
and boundary conditions,
\begin{subequations} 
\begin{align} 
&\vn\times\vE_a={\bf 0}, \  
\vn\cdot\vH_a=0, \ 
\vn\times\vF_a={\bf 0}, \ 
\psi_a=0 \ \ \ {\rm on \ S}, \\
&\vn\times\vH_a={\bf 0}, \ 
\vn\cdot\vE_a=0, \ 
\vn\times\vF_a={\bf 0}, \ 
\psi_a=0 \ \ \ {\rm on \ S'}
\,, 
\end{align}\label{eq-boundary}
\end{subequations} 
where $\vn$ is the outer normal to the surface, and $S$ and $S'$ are the conductive and insulating surfaces, respectively. We can also expand $\nabla\times\vE$, $\nabla\times\vH$, and $\nabla\cdot\vE$ in terms of the functions $\vE_a$, $\vH_a$, $\vF_a$, and $\psi_a$ as \cite{sla46}, 
\begin{subequations} 
\begin{align} 
\nabla\times\vE&=\sum_a\vH_a\int(\nabla\times\vE)\cdot\vH_adv \notag \\
&=\sum_a\vH_a\left[k_a\int\vE\cdot\vE_adv
+\int_S(\vn\times\vE)\cdot\vH_ada\right], \\
\nabla\times\vH&=\sum_a\vE_a\int(\nabla\times\vH)\cdot\vE_adv \notag \\
&=\sum_a\vE_a\left[k_a\int\vH\cdot\vH_adv
+\int_{S'}(\vn\times\vH)\cdot\vE_ada\right], \\
\nabla\cdot\vE&=\sum_a\psi_a\int(\nabla\cdot\vE)\psi_adv 
=-\sum_a\psi_ak_a\int\vE\cdot\vF_adv
\,.
\end{align}
\label{eq-expansion2}
\end{subequations} 
Substituting Eqs.~\eqref{eq-expansion1} and \eqref{eq-expansion2} into Maxwell's equations, we obtain
\begin{subequations} 
\begin{align} 
&k_a\int\vE\cdot\vE_adv+\frac{1}{c}\frac{d}{dt}\int\vH\cdot\vH_adv \notag \\
& \ \ \ \ \ =-\int_S(\vn\times\vE)\cdot\vH_ada, \label{eq-Faraday} \\
&k_a\int\vH\cdot\vH_adv-\frac{1}{c}\frac{d}{dt}\int\vE\cdot\vE_adv \notag \\ 
& \ \ \ \ \ =\frac{4\pi}{c}\int\vj\cdot\vE_adv 
-\int_{S'}(\vn\times\vH)\cdot\vE_ada, \label{eq-Ampere} \\
&-\frac{1}{c}\frac{d}{dt}\int\vE\cdot\vF_adv 
=\frac{4\pi}{c}\int\vj\cdot\vF_adv, \label{eq-Ampere2} \\
&-k_a\int\vE\cdot\vF_adv 
=4\pi\int\rho\psi_adv. \label{eq-Gauss} 
\end{align} 
\end{subequations} 
We also obtain separate equations for $\int\vE\cdot\vE_adv$ and $\int\vH\cdot\vH_adv$ from Eqs. (\ref{eq-Faraday}) and (\ref{eq-Ampere}) as 
\begin{subequations} 
\begin{align} 
\frac{1}{c^2}&\frac{d^2}{dt^2}\int\vE\cdot\vE_adv 
+k_a^2\int\vE\cdot\vE_adv \notag \\
=&-\frac{1}{c}\frac{d}{dt}
\left[\frac{4\pi}{c}\int\vj\cdot\vE_adv-\int_{S'}(\vn\times\vH)\cdot\vE_ada\right] \notag \\
&-k_a\int_S(\vn\times\vE)\cdot\vH_ada, \label{eq-cavityE} \\
\frac{1}{c^2}&\frac{d^2}{dt^2}\int\vH\cdot\vH_adv 
+k_a^2\int\vH\cdot\vH_adv \notag \\
=&k_a\left[\frac{4\pi}{c}\int\vj\cdot\vE_adv-\int_{S'}(\vn\times\vH)\cdot\vE_ada\right] \notag 
\\
&-\frac{1}{c}\frac{d}{dt}\int_S(\vn\times\vE)\cdot\vH_ada. 
\end{align}\label{eq-cavity}% 
\end{subequations} 
These are the key equations for the EM fields in a hollow cavity. We also obtain the continuity equation from Eqs.~\eqref{eq-Ampere2} and \eqref{eq-Gauss}.

First consider an ideal resonant cavity without damping, and assume that the current density is zero in the cavity.  Then, the right-hand sides of Eq.~\eqref{eq-cavity} are zero, since $\vE$ and $\vH$ satisfy $\vn\times\vE={\bf 0}$ on $S$ and $\vn\times\vH={\bf 0}$ on $S'$ as well as $\vE_a$ and $\vH_a$ in Eq.~\eqref{eq-boundary}, respectively. Therefore, the solutions for Eq.~\eqref{eq-cavity} are given by
\begin{subequations} 
\begin{align} 
\int\vE\cdot\vE_adv=C{\rm e}^{-i\omega_at}, \
&\int\vH\cdot\vH_adv=C'{\rm e}^{-i\omega_at}, \\ 
\omega_a&=k_ac, 
\end{align}
\end{subequations} 
where $C$ and $C'$ are constants. We also obtain $\int\vE\cdot\vE_adv/\int\vH\cdot\vH_adv$ from Eq.~\eqref{eq-Faraday} or \eqref{eq-Ampere} as
\begin{align} 
\frac{\displaystyle\int\vE\cdot\vE_adv}{\displaystyle\int\vH\cdot\vH_adv}
=i
\,.
\label{eq-C/C'}
\end{align}
Thus, $\omega_a$ is the angular frequency of $a$th mode in an ideal resonant cavity with $k_a$, which is the propagation constant of the $a$th mode obtained from Eqs.~\eqref{eq-wave_eq} and \eqref{eq-boundary}.

We next consider a resonant cavity with small damping and assume that $\int\vE\cdot\vE_adv $ varies as ${\rm e}^{-i\omega t}$. Substituting $k_a=\omega_a/c$ into Eq.~\eqref{eq-cavityE} we obtain,
\begin{align} 
\bigg(\frac{\omega}{\omega_a}&-\frac{\omega_a}{\omega}\bigg)
\int\vE\cdot\vE_adv \notag \\
=&-i\frac{c}{\omega_a}
\left[\frac{4\pi}{c}\int\vj\cdot\vE_adv-\int_{S'}(\vn\times\vH)\cdot\vE_ada\right] \notag \\
&+\frac{c}{\omega}\int_S(\vn\times\vE)\cdot\vH_ada. \label{eq-cavityE2}
\end{align}
Assuming that the form of the complex angular frequency in a resonant cavity with small damping to be $\omega=\omega_a+\Delta\omega_a-i(\omega_a/2Q)$ in the limit of $Q\gg1$ and $\omega_a\gg\Delta\omega_a$, we then obtain  
\begin{align} 
\left(\frac{\omega}{\omega_a}-\frac{\omega_a}{\omega}\right)
=2\frac{\Delta\omega_a}{\omega_a}-\frac{i}{Q}, \label{eq-frequency}
\end{align}
where $Q$ and $\Delta\omega_a$ are the quality factor and the frequency shift from an ideal cavity~\cite{jac98}.
Substituting Eq.~\eqref{eq-frequency} into Eq.~\eqref{eq-cavityE2}, and replacing $\omega$ to $\omega_a$ in the right-hand side of Eq.~\eqref{eq-cavityE2}, we obtain Eq.~\eqref{eq-Q_and_Delta_omega}.

\medskip

%-------------------------------------------------------------------
\section{Perturbative correction to $\delta f$ and $Q$ near $T_c$}

To derive expression for the frequency shift very near $T_c$ we carry out a perturbative expansion of Eq.~\eqref{eq-sigma} for $\sigma(\omega)$ to second order in $\Delta(T)$,
%
%--------------------------------------------------------------------
\onecolumngrid
\begin{align} 
\sigma& = \sigma_n + \frac{\sigma_D}{i\omega\tau}
\int_{-\infty}^\infty \frac{d\varepsilon}{4\pi i}
\bigg\{
%reterded part
\tanh\left(\frac{\varepsilon-\omega/2}{2T}\right) 
\frac{\pi}{-2i(\varepsilon+i\eta)+1/\tau} \notag \\
&\times\bigg[
\frac{\Delta^{2}}{(\varepsilon+\omega/2+i\eta)(\varepsilon-\omega/2+i\eta)}
+\frac{\Delta^{2}}{2(\varepsilon-\omega/2+i\eta)^2}
+\frac{\Delta^{2}}{2(\varepsilon+\omega/2+i\eta)^2}
\bigg] \notag \\
%advanced part
-&\tanh\left(\frac{\varepsilon+\omega/2}{2T}\right) 
\frac{\pi}{2i(\varepsilon-i\eta)+1/\tau} \notag \\
&\times\bigg[
\frac{\Delta^{2}}{(\varepsilon+\omega/2-i\eta)(\varepsilon-\omega/2-i\eta)}
+\frac{\Delta^{2}}{2(\varepsilon-\omega/2-i\eta)^2}
+\frac{\Delta^{2}}{2(\varepsilon+\omega/2-i\eta)^2}
\bigg] \notag \\
%anomalous part
+&\bigg[\tanh\left(\frac{\varepsilon-\omega/2}{2T}\right)
-\tanh\left(\frac{\varepsilon+\omega/2}{2T}\right)\bigg] 
\frac{\pi}{-i\omega+1/\tau} \notag \\
&\times\bigg[
\frac{\Delta^{2}}{(\varepsilon+\omega/2+i\eta)(\varepsilon-\omega/2-i\eta)}
+\frac{\Delta^{2}}{2(\varepsilon-\omega/2-i\eta)^2}
+\frac{\Delta^{2}}{2(\varepsilon+\omega/2+i\eta)^2}
\bigg] 
\bigg\}. 
\end{align}
Performing the energy integral using the residue theorem choosing integration paths such that the integrands are analytic, and replacing $T$ with $T_c$ except in $\Delta$, we obtain the real and imaginary parts of conductivity as
$\sigma_1=\sigma_{1n}+\sigma_1^{(2)}$ and $\sigma_2=\sigma_{2n}+\sigma_2^{(2)}$, with
%--------------------------------------------------------------------
\begin{align} 
\sigma_1^{(2)}
&=-2\pi T_c\frac{\sigma_D}{\tau}\Delta^2 \sum_{n=0}^\infty
\bigg\{
\frac{2\varepsilon_n+1/\tau}{(2\varepsilon_n+1/\tau)^2+\omega^2} 
\bigg[
\frac{1}{\varepsilon_n(\varepsilon_n^2+\omega^2)}
+\frac{\varepsilon_n}{(\varepsilon_n^2+\omega^2)^2}
\bigg] \notag \\
&+\frac{1}{(2\varepsilon_n+1/\tau)^2+\omega^2} 
\bigg[
\frac{1}{\varepsilon_n^2+\omega^2}
+\frac{\varepsilon_n^2-\omega^2}{2(\varepsilon_n^2+\omega^2)^2}
+\frac{1}{2\varepsilon_n^2}
\bigg]
\notag \\
&+\frac{1/\tau}{1/\tau^2+\omega^2} 
\bigg[
\frac{1}{\varepsilon_n(\varepsilon_n^2+\omega^2)}
+\frac{\varepsilon_n}{(\varepsilon_n^2+\omega^2)^2}
\bigg] 
\notag \\
&+\frac{1}{1/\tau^2+\omega^2} 
\bigg[
\frac{1}{\varepsilon_n^2+\omega^2}
+\frac{\varepsilon_n^2-\omega^2}{2(\varepsilon_n^2+\omega^2)^2}
-\frac{1}{2\varepsilon_n^2}
\bigg] 
\bigg\}, 
\label{eq-sigma1_and_sigma2_near_Tc}
\end{align} 
\begin{align} 
\sigma_2^{(2)}
&=
\frac{2\pi T_c}{\omega}\frac{\sigma_D}{\tau}\Delta^2 \sum_{n=0}^\infty 
\bigg\{
\frac{2\varepsilon_n+1/\tau}{(2\varepsilon_n+1/\tau)^2+\omega^2} 
\bigg[
\frac{1}{\varepsilon_n^2+\omega^2}
+\frac{\varepsilon_n^2-\omega^2}{2(\varepsilon_n^2+\omega^2)^2}
+\frac{1}{2\varepsilon_n^2}
\bigg] \notag \\
&-\frac{\omega^2}{(2\varepsilon_n+1/\tau)^2+\omega^2} 
\bigg[
\frac{1}{\varepsilon_n(\varepsilon_n^2+\omega^2)}
+\frac{\varepsilon_n}{(\varepsilon_n^2+\omega^2)^2}
\bigg] \notag \\
&+\frac{1/\tau}{1/\tau^2+\omega^2} 
\bigg[
\frac{1}{\varepsilon_n^2+\omega^2}
+\frac{\varepsilon_n^2-\omega^2}{2(\varepsilon_n^2+\omega^2)^2}
-\frac{1}{2\varepsilon_n^2}
\bigg] \notag \\
&-\frac{\omega^2}{1/\tau^2+\omega^2} 
\bigg[
\frac{1}{\varepsilon_n(\varepsilon_n^2+\omega^2)}
+\frac{\varepsilon_n}{(\varepsilon_n^2+\omega^2)^2}
\bigg] 
\bigg\}, 
\label{eq-sigma_near_Tc}%
\end{align}
\twocolumngrid
\noindent where $\varepsilon_n=(2n+1)\pi T_c$ is the Matsubara energy at $T=T_c$. Using Eqs.~\eqref{eq-Zs} and \eqref{eq-sigma_near_Tc}, and $\sigma_{2n}\approx 0$, the surface resistance and reactance to the second order in $\Delta$ are given by 
\begin{subequations}
\begin{align} 
\frac{R_s}{R_n}&=1-\frac{\sigma_1^{(2)}+\sigma_2^{(2)}}{2\sigma_{1n}}, \\
\frac{X_s}{R_n}&=1-\frac{\sigma_1^{(2)}-\sigma_2^{(2)}}{2\sigma_{1n}}.  
\end{align}\label{eq-Zs_near_Tc}%
\end{subequations}
We finally obtain the frequency shift and quality factor very near $T_c$ from Eqs.~\eqref{eq-Delta_f},~\eqref{eq-Q} and~\eqref{eq-Zs_near_Tc}
\begin{subequations}
\begin{align} 
\delta f&=\frac{fR_n}{4G}\frac{\sigma_1^{(2)}-\sigma_2^{(2)}}{\sigma_{1n}}, 
\label{eq-Delta_f_near_Tc} \\ 
Q&=\frac{G}{R_n}+\frac{G}{2R_n}\frac{\sigma_1^{(2)}+\sigma_2^{(2)}}{\sigma_{1n}}. 
\end{align}
\end{subequations}
Thus, as seen in Eq. (\ref{eq-Delta_f_near_Tc}), we find an anomaly where the frequency shift is negative in a superconductor near $T_{c}$ when $\sigma_2^{(2)}$ is larger than $\sigma_1^{(2)}$.

%-------------------------------------------------------------------

\section{Inhomogenous Disorder}
\label{appendix-Inhomogeneous_Disorder}

We assume that $T_c$ is Gaussian distributed, and we replace $\Delta_0$ with $\Delta_0=\pi e^{-\gamma}T_c^{\text{av}}$, where $T_c^{\text{av}}$ is the average value of $T_c$ for the cavity. 
The distribution of $T_c$ is then given by~\footnote{We choose Gaussians for the distributions of $T_c$ and $1/\tau$ defined in Eqs.~\eqref{eq-GaussianTc} and~\eqref{eq-Gaussianrho}. In all cases of interest the variance $\mu_2$ is small compared to the mean, $\mu$. Thus, we can extend the domain to $(-\infty,+\infty)$ and compute the prefactors analytically as shown in Eqs~\eqref{eq-GaussianTc} and~\eqref{eq-Gaussianrho}.} 
\begin{align} 
\rho(T_c)
=\frac{1}{\sqrt{2\pi\mu_2}}{\rm e}^{-\frac{(T_c-\mu)^2}{2\mu_2}}, 
\label{eq-GaussianTc}
\end{align}
where $\mu$ and $\sqrt{\mu_2}$ are the average and variance, respectively. 

\begin{figure*}[t]
\includegraphics[width=\linewidth]{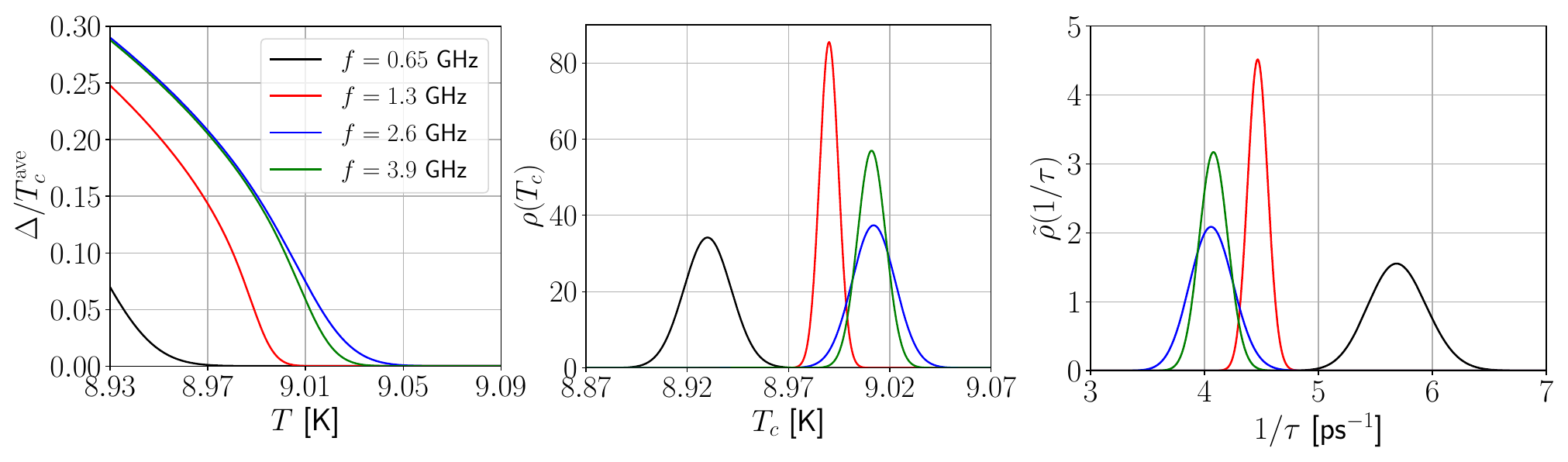}
\caption{Gap energy $\Delta$ near $T_c$ (left), Gaussian distributions of the transition temperature $\rho(T_c)$ (center), and probability density functions of the relaxation time inverse $\tilde{\rho}(1/\tau)$ (right), for the Nb SRF cavities with each different resonance frequency, $f=0.65\,\mbox{GHz}$ (black), $1.3\,\mbox{GHz}$ (red), $2.6\,\mbox{GHz}$ (blue), and $3.9\,\mbox{GHz}$ (green).
}
\label{fig-inhomogeneity}
\end{figure*}

We interpret the spread in $T_c$ in terms of inhomogeneity of the disorder, i.e. $1/\tau$. Thus, we infer the probability distribution function for the scattering rate from the Gaussian distribution from the transition temperature,
We obtain the probability distribution for $1/\tau$ from Eq.\eqref{eq-GaussianTc} and the normalization condition, 
\begin{align} 
\int_{-\infty}^{\infty}dT_c\rho(T_c)=\int_{-\infty}^{\infty}d\left(\frac{1}{\tau}\right)\tilde{
\rho}(1/\tau)=1
\,, 
\label{eq-Gaussianrho}
\end{align}
as 
\begin{align} 
\tilde{\rho}(1/\tau)=-\frac{dT_c(1/\tau)}{d(1/\tau)}\frac{1}{\sqrt{2\pi\mu_2}}
{\rm e}^{-\frac{[T_c(1/\tau)-\mu]^2}{2\mu_2}}.
\label{eq-probability_density_tau}
\end{align}
Note that $\mu$ and $\mu_2$ are the same as those in Eq. (\ref{eq-GaussianTc}). 
Function $T_c(1/\tau)$ is calculated by solving Eq.~\eqref{eq-Tc}, 
and $dT_c(1/\tau)/d(1/\tau)$ is obtained by differentiating Eq.~\eqref{eq-Tc} to obtain,
\begin{align} 
&\frac{dT_c(1/\tau)}{d(1/\tau)} =-\frac{T_c}{1/\tau} \notag \\
&\times
\frac{\displaystyle\frac{\cA}{2}\frac{1/\tau}{2\pi T_{c0}}
\sum_{n=0}^\infty\frac{1}{\left(n+\frac{1}{2}+\frac{1}{2}\frac{1/\tau}{2\pi T_c}\right)^2}}
{\displaystyle\frac{T_c}{T_{c0}}-\frac{\cA}{2}\frac{1/\tau}{2\pi T_{{c0}}}
\sum_{n=0}^\infty\frac{1}{\left(n+\frac{1}{2}+\frac{1}{2}\frac{1/\tau}{2\pi T_{c}}\right)^2}}. 
\end{align}
Figure~\ref{fig-inhomogeneity} shows the temperature dependence of the disorder-averaged gap near $T_c$, Gaussian distributions of $T_c$, and probability distribution function of $1/\tau$ for the four N-doped Nb SRF cavities.

%----------------------------------------------------------
%----------------------------------------------------------

%---------------------------------------------------------------------
%\bibliography{CM,SC,SRF,QFS,Books}
%---------------------------------------------------------------------
%
\end{document}